\documentclass[twocolumn,showpacs,preprintnumbers,amsmath,amssymb]{revtex4}
\usepackage{graphicx}

\def\lsim{\mathrel{\rlap{\lower4pt\hbox{\hskip1pt$\sim$}}
    \raise1pt\hbox{$<$}}}                
\def\gsim{\mathrel{\rlap{\lower4pt\hbox{\hskip1pt$\sim$}}
    \raise1pt\hbox{$>$}}}                

\begin{document}
\title{The Large Scale Curvature of Networks}

\author{Onuttom Narayan$^1$ and Iraj Saniee$^2$}
\affiliation{$^1$ Department of Physics, University of California, Santa Cruz, CA 95064}
\affiliation{$^2$ Mathematics of Networks Department, Bell Laboratories,
Alcatel-Lucent, 600 Mountain Avenue, Murray Hill, NJ 07974}
\date{\today}
\begin{abstract}
Understanding key structural properties of large scale networks are
crucial for analyzing and optimizing their performance, and improving
their reliability and security.  Here we show that these networks possess
a previously unnoticed feature, global curvature, which we argue has a
major impact on core congestion: the load at the core of a network
with $N$ nodes scales as $N^2$ as compared to $N^{1.5}$ for a flat
network. We substantiate this claim through analysis of a collection
of real data networks across the globe as measured and documented by
previous researchers.
\end{abstract}
\maketitle

Large-scale data networks form the infrastructure for contemporary global
communications. Increasingly, the trend in these networks is towards
converged services over the Internet protocol, dynamic and automatic
reconfigureability, and flatter architecture for fast service creation
and survivability. In such a large and fast changing environment, there
is a need for identifying key structural properties that affect their
performance, reliability and security and which provide efficient and
scalable models to estimate these metrics reliably.

Recent models of networks have focused on features such as
their `small world' property~\cite{comm1,erdos,sw1} or power law
degree distributions~\cite{ab1,krapivsky,dorog}.  There has been
evidence for power-law degree distributions in data networks at the
IP layer~\cite{faloutsos}, for the worldwide web~\cite{ab1}, and
even for the virtual network of social connections~\cite{barabasi2},
but are found not to exist for physical networks such as electrical
grids~\cite{sw1,newman} and some biological networks~\cite{newman,white}.
Although these features are interesting and important, the impact of
intrinsic geometrical and topological features of large-scale networks
on performance, reliability and security is of much greater importance.
Intuitively, it is known that traffic between nodes tends to go through
a relatively small core of the network~\cite{fn2} as if the shortest
path between them is curved inwards. It has been suggested that this
property may be due to {\it global curvature} of the network~\cite{yuliy}.

In this paper, we define the global (negative) curvature for finite
networks and demonstrate its existence at the IP layer by examining
topologies of numerous publicly available networks~\cite{rocketfuel}.  A recent
report~\cite{kriou3}, also refers to curvature as a possible cause of
some key observations about networks at the IP layer. However, these
authors {\it assume\/} negative curvature, and construct a model with
a few extra simple assumptions that shares various features with real
networks such as a power law degree distribution. By contrast, we {\it
demonstrate\/} negative curvature through direct measurement.

Turning to the impact of negative curvature, we focus on the load (also
referred to as the betweenness centrality), as defined by assuming
unit traffic between each pair of nodes in the network with shortest
path routing, and calculating the traffic through each node. (This is
{\it not\/} the actual time-variable demand that is routed through
nodes and links at the IP layer.) We show that network curvature or
$\delta$-hyperbolicity~\cite{gromov} implies that the load at the core
of the network scales with the number of nodes $N$ as $\sim N^2,$ which
is faster than the $\sim N^{1.5}$ scaling for flat networks.  Thus core
congestion is worse in hyperbolic networks, and geodesic routing achieved
with greedy algorithms on hyperbolic networks~\cite{kriou3} is actually
problematic. Previous work~\cite{gkk,gjkk,barth} has considered the load
as a function of node degree for fixed $N,$ which we have also examined
separately~\cite{ns}.

\begin{figure}[htb]
\begin{center}
\includegraphics[width = \columnwidth]{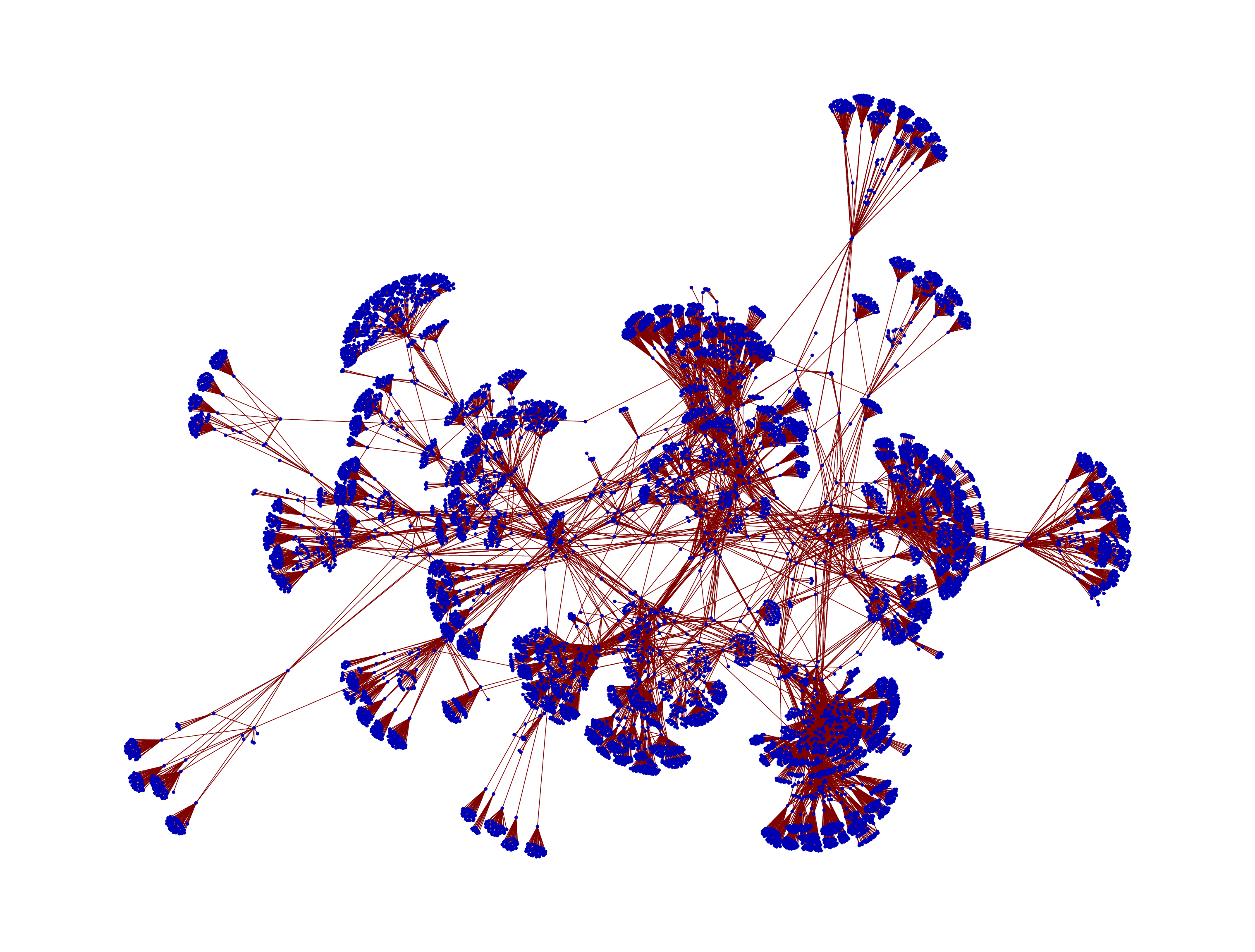}
\caption{
A rendering of the graph for the network 7018(AT\&T).}
\label{fig:tt}
\end{center}
\end{figure}
Negative curvature of a geodesic metric space is defined by
Gromov~\cite{gromov} in terms of the `$\delta$-Thin Triangle Condition'.
For a graph, an appropriate metric can be used. For any three nodes
$(ijk),$ the geodesics $g_{ij}, g_{jk}$ and $g_{ki}$ of lengths $d_{ij},
d_{jk}$ and $d_{ki}$ are constructed.  A fourth node $m$ is chosen,
and the shortest distance between $m$ and all the nodes on $(ij)$ is
defined as $d(m;ij).$ The distance $D(m;ijk)$ is defined as the maximum
of $d(m;ij),d(m;jk)$ and $d(m;ki).$ Then if
\begin{equation}
\max_{(ijk)}\min_m D(m;ijk) = \delta
\label{eq:gromov_def}
\end{equation}
is finite, the (infinite) graph is said to have negative or hyperbolic
curvature. Other definitions of curvature count the triangles (or other
polygons) that meet at each vertex\cite{gauss-bonnet}, but these define
a local, not global, curvature and can be argued to be unrelated to the
global performance of networks.

For a finite graph, Eq.(\ref{eq:gromov_def}) is trivially finite and the
Gromov curvature has to be modified.  We introduce the concept of the
``curvature plot" of a network: for every triangle $\Delta = (ijk)$
we plot $\delta_\Delta$ vs $L_\Delta$ where
\begin{eqnarray}
\delta_\Delta &=& \min_m D(m;ijk)\nonumber\\
L_\Delta &=& \min [d(ij), d(jk), d(ki)].
\label{eq:finitegraph}
\end{eqnarray}
This yields $P_L(\delta),$ the probability distribution for $\delta$
at fixed $L.$ If the peak of $P_L(\delta)$ is at $\delta=\delta_p(L),$
the network is flat (negatively curved) if $\delta_p(L)$ increases
linearly (sublinearly) with $L$~\cite{fn3}.  Since we use the peak of
the distribution instead of the maximum as in Eq.(\ref{eq:gromov_def}),
statistical sampling of triangles is sufficient.

\begin{figure}
\begin{center}
\includegraphics[width = \columnwidth]{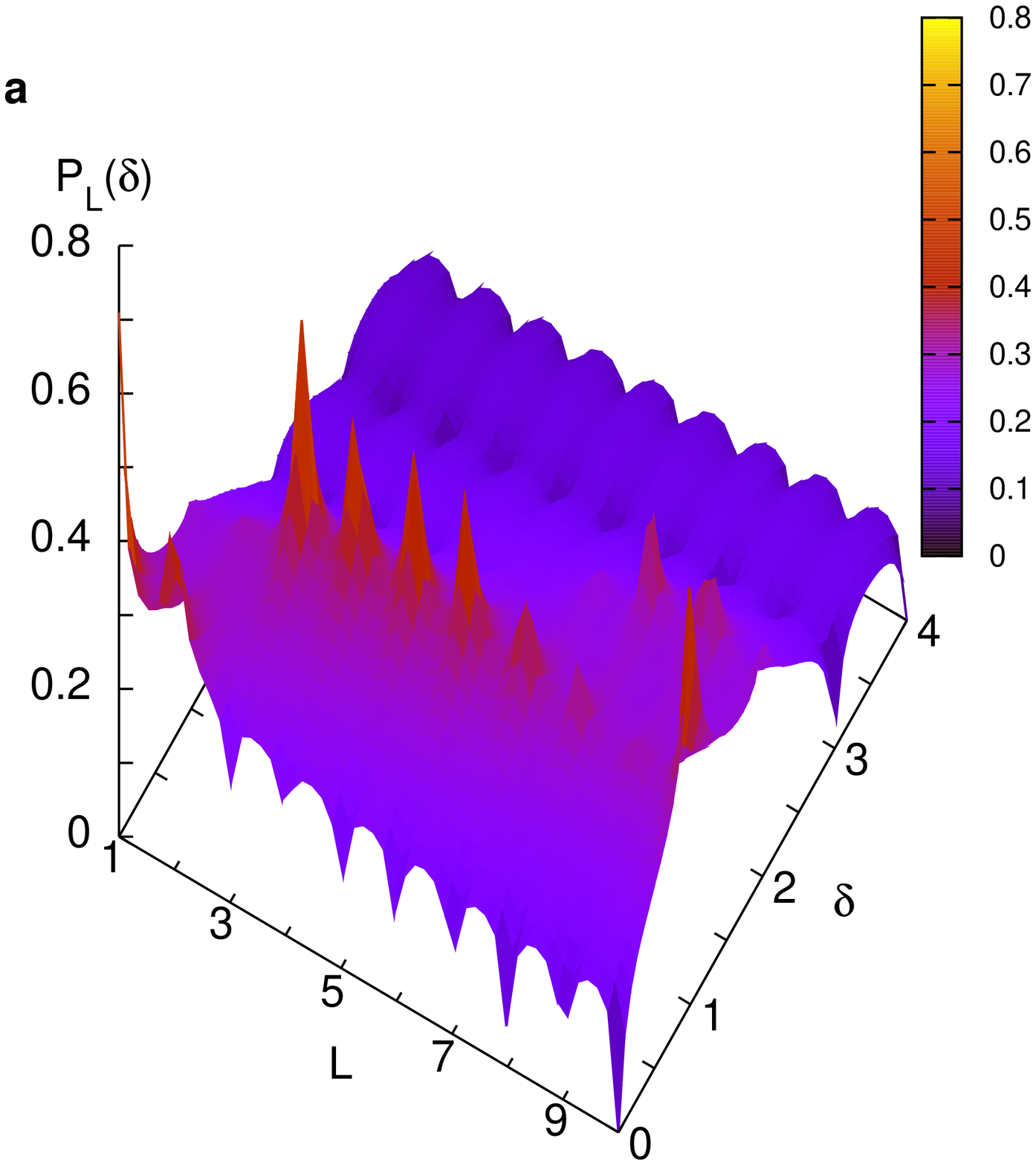}
\includegraphics[width = \columnwidth]{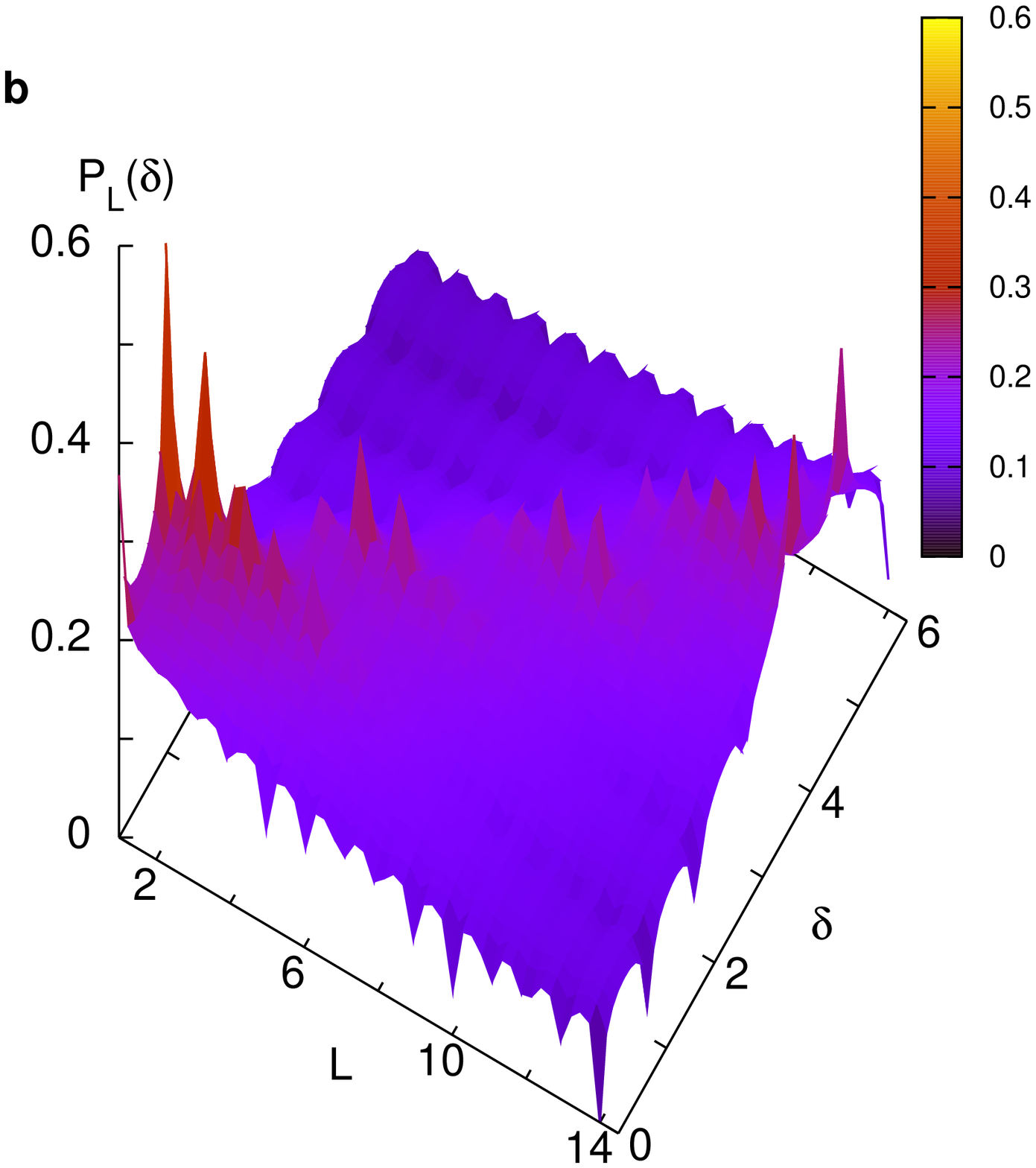}
\caption{(a) Probability $P_L(\delta)$ for randomly 
chosen triangles whose shortest side is $L$ to have a given $\delta$ as
defined in Eq.(\ref{eq:finitegraph}) for the network 7018(AT\&T network)
which has 10152 nodes and 14319 links and diameter 12.  The quantities
$\delta$ and $L$ are restricted to integers, and the smooth plot is by
interpolation. (b) Similar to (a), for a (flat) triangular lattice with
469 nodes and 1260 links.  (The smaller number of nodes is sufficient for
comparing with (a) since the range for $L$ is large due to the absence
of the small world effect.)}
\label{fig:gromovplot}
\end{center}
\end{figure}
Figure~\ref{fig:gromovplot} shows the curvature plot for network
7018(AT\&T) from the Rocketfuel database~\cite{rocketfuel} (see
Figure~\ref{fig:tt}). The metric used is the `hop metric', where each
edge of the graph has unit length.  This is a common metric that best
illustrates the geometrical properties of the graph, including the `small
world' property~\cite{fn1}.  The networks in this database are at the IP
layer and describe the IP port to IP port connectivity of the network.
A sharp ridge is seen along the curve $\delta_p(L).$ The ridge is a
straight line through the origin for the triangular lattice but bends
over parallel to the $L$-axis for the 7018 network ($P_L(\delta)$ is
{\it zero} for $\delta > 3$ for all $L$, though the diameter of the
network is 12).  For all the networks in the database, we have verified
that the measured $\delta$'s do not exceed 3, even though the network
diameters range from  12 to 14 (with the exception of 4755/VSNL whose
diameter is 6, but whose ratio diameter/$\delta$ is even bigger, 6).
The ratio of 3/12 or 25\% is comfortably within the theoretical bound
for scaled hyperbolic graphs\cite{jonck}.

\begin{figure}[htb]
\begin{center}
\includegraphics[angle= 270, width =\columnwidth]{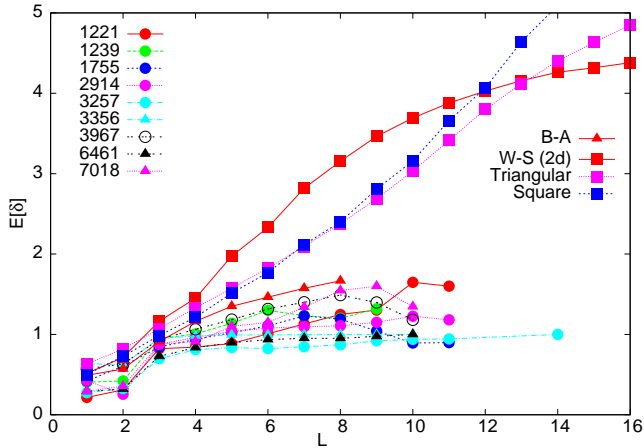}
\caption{The average $\delta$ as a function of 
$L,$ $E[\delta](L),$ for the 10 IP-layer networks studied here, and
for the Barabasi-Albert model with $k=2$ and $N=10000$ (11th curve)
and the hyperbolic grid $X_{3,7}$ (12th curve).  On the other hand,
a Watts-Strogatz type model on a square lattice with $N=6400,$ open
boundary conditions and $5\%$ extra random connections (13th curve)
and two flat grids (the triangular lattice with diameter 29  and the
square lattice with diameter 154) are also shown.}
\label{fig:gromovgraph}
\end{center}
\end{figure}

As another manifestation of the curvature, Figure~\ref{fig:gromovgraph}
shows the average $\delta$ for each $L,$ $E[\delta](L),$ for all ten
networks in the Rocketfuel database. The plots saturate for relatively
small $L.$ The figure also shows $E[\delta](L)$ for the Barabasi-Albert
model~\cite{ab1} and a Watts-Strogatz type model~\cite{sw1}; although
both of these models exhibit small world behavior, we see that only the
first has negative curvature as defined in Eq.(\ref{eq:gromov_def}). The
plot for the Watts-Strogatz graph shows signs of saturation for large
$L$, but the size of this graph was chosen so that it was already well
in the small world regime~\cite{sw1,newman}.

\begin{figure}[htb]
\begin{center}
\includegraphics[angle= 270, width = \columnwidth]{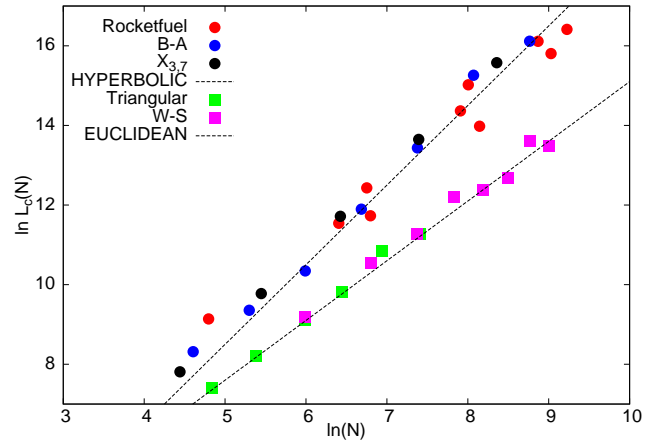}
\caption{Plot of the maximum load $L_c(N)$ for each 
network in the Rocketfuel database as a function of the number of nodes
$N$ in the network.  Also shown are the maximum load for the hyperbolic
grid $X_{3,7},$ the Barabasi-Albert model with $k=2,$ the Watts-Strogatz
model and a triangular lattice, for various $N.$ The dashed lines have
slopes of 2.0 and 1.5, corresponding to the hyperbolic and Euclidean
cases respectively.}
\label{fig:load}
\end{center}
\end{figure}

Turning to the performance implications of hyperbolic curvature, the
simplest graphs with (constant) negative curvature are the hyperbolic
grids $X_{p,q}$ consisting of $q$ regular $p$-gons at each vertex when
$(p-2)(q-2)>4$.\cite{fn5} (When $(p-2)(q-2)=4$, the graph is flat.)
We construct finite hyperbolic grids by truncating to $n$ hops from the
center. The number of nodes $N$ in the graph increases exponentially
as $n$ is increased.  With unit demand between all node pairs and the
traffic between two nodes traveling along a geodesic connecting them
(evenly distributed over all geodesics in case of ties), we have verified
numerically that the load at the center of the graph scales with the
number of nodes $N$ in the graph as
\begin{equation}
L_c(N)\sim N^2.
\label{eq:coreload}
\end{equation}
The same result can be obtained analytically for the continuum
Poincar\'{e} disk truncated to a radius $r < 1,$ converted to a graph
by introducing a uniform distribution of nodes with each node connected
to its neighbors.\cite{fn4} By contrast, it is not hard to verify that
$L_c(N)\sim N^{1.5}$ for a Euclidean graph.  Physically, this is because
the traffic from the $\sim N$ nodes on the left of a Euclidean lattice to
the $\sim N$ nodes on the right flows through the center across a line of
length $\sim\sqrt N,$ whereas for a hyperbolic graph it is pulled inwards
and flows within an $O(1)$ distance from the center.  
Figure~\ref{fig:load} shows
the load at the node with the highest load versus $N$ for all the networks
in the Rocketfuel database, demonstrating $\sim N^2$ scaling. The figure
also shows results for the Barabasi-Albert and Watts-Strogatz models;
we see that the first shows $\sim N^2$ scaling but the second does not,
confirming our earlier conclusion that the latter is a poorer fit to
Internet-type large-scale networks.

\begin{figure}[htb]
\begin{center}
\includegraphics[width = 3in]{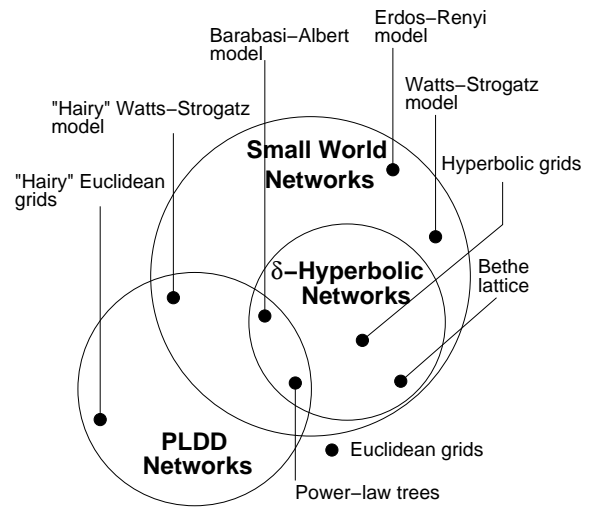}
\caption{{\bf Taxonomy:} Taxonomy of key characteristics of networks
and their overlaps in a schematic diagram. ``Hairy'' as used in this
figure, refers to the simple mechanism of making a grid power-law by
adding to each node a set of singly-connected nodes (hairs) whose number
is drawn from any desired power-law distribution. PLDD refers to power
law degree distributions.}
\label{fig:taxonomy}
\end{center}
\end{figure}

There are two points worth noting. First, one might wonder whether the
concentration of geodesics and load near the center is trivial because
the networks we have studied are almost simple trees.  However, the
ratio of the number of edges to nodes in these networks ranges from
1.27 to 2.72, showing that they are far from being trees. Second, as
the example of hyperbolic grids demonstrate, one can construct graphs
where every node has the same degree, but which exhibit the `small
world' property {\em and} show $N^2$ scaling of load.  Thus although the
networks we have studied do seem to have power-law degree distributions,
hyperbolicity is a nontrivial and general property that is distinct
from their degree distribution and --- based on the $\sim N^2$ scaling
of the previous paragraph --- can significantly impact performance.
Figure~\ref{fig:taxonomy} summarizes the relationship between several key
characteristics discussed in the literature in the context of large-scale
complex networks.  We observe that hyperbolicity entails small world
behavior, a fundamental property of networks.

Our results suggest that, counter-balancing the positive benefits of
hyperbolicity such as the small world property, core congestion is a
structural {\it problem} due such hyperbolicity that grows more acute as
the network grows in size.  As long as routing protocols use geodesics
in one form or another, whether in intra-domain, inter-domain or other
forms of routing, congestion is a natural consequence of this intrinsic
structural feature of networks.  Using `$(1+\epsilon)$ routing', in which
traffic between nodes is {\it not\/} routed along the geodesic(s) between
them but is deliberately sent on slightly longer paths, would in fact
alleviate core congestion.  This is a phenomenon familiar from vehicular
traffic: the shortest routes using expressways can become so overcrowded
that indirect and longer paths through backroads become faster. 

This work was funded by AFOSR grant FA9550-08-1-0064.


\begin{thebibliography}{99}
\bibitem{comm1}
This is the same as the shortest path length when each edge is given weight 1.
\bibitem{fn2}
In this context, the `network core' is the set of nodes that are at
the intersection of the majority of geodesics (shortest paths), while
the `network edge' is the set where this property is minimal.
\bibitem{fn3}
By choosing $m$ to be on the sides of the triangle, it is easy to verify
that $\delta_\Delta \leq L_\Delta/2,$ so that positive curvature cannot
be seen with this test or Eq.(\ref{eq:gromov_def}).
\bibitem{fn1}
While routing in the
Internet can involve link metrics other than the hop metric, and careful
sizing of links and assignment of such metrics can avoid congestion in
prespecified hot spots in the network, such measures often create hot
spots elsewhere.\cite{internetroutingproblem}.
\bibitem{fn4}
This model is applicable to a hyperbolic network if the spacing between 
nodes is small compared to its radius of curvature.
\end{thebibliography}
\end{document}